# AN INVESTIGATION AND INSIGHT INTO TERRORISM IN NIGERIA

**Aamo Iorliam[1], Raymond U. Dugeri[2], Beatrice O. Akumba[3], Samera Otor[4], and Yahaya I. Shehu[5]**
**[1,2,3,4] Department of Mathematics & Computer Science, BSU, Makurdi, Nigeria**
**[5]Shehu Shagari College of Education, Sokoto, Nigeria**
Corresponding author: Iorliam, A. (aamoiorliam@gmail.com).

## ABSTRACT

*Terrorism is one of the most life-challenging threats facing humanity worldwide. The activities of terrorist organizations threaten peace, disrupts progress, and halt the development of any nation. Terrorist activities in Nigeria in the last decades have negatively affected economic growth and have drastically reduced the possibilities of foreign investments in Nigeria. In this paper, statistical and inferential insights are applied to the terrorist activities in Nigeria from 1970 to 2019. Using the Global Terrorism Database (GTD), insights are made on the occurrences of terrorist attacks, the localities of the target, and the successful and unsuccessful rates of such attacks. The Apriori algorithm is also used in this paper to draw hidden patterns from the GTD to aid in generating strong rules through database mining, resulting in relevant insights. This understanding of terrorist activities will provide security agencies with the needed information to be one step ahead of terrorists in making the right decisions targeted at curbing terrorism in Nigeria.*

# 1.0 INTRODUCTION

Terrorism has become so rampant recently in Nigeria due to the porous nature of the land borders and lack of proper digital surveillance of the Nigerian digital space (Yakubu, and Idehen, 2021; Iorliam, 2019). Terrorists cause a lot of unrest, fear, and destruction in Nigeria and the entire world.

The destruction of lives and properties has been the most detrimental effect of terrorism in Nigeria. This act has also heightened fear, reduced life expectancy, hindered infrastructural development, and negatively tainted Nigeria's international image (Agu, 2018).

According to the Global Conflict Tracker (2021), terrorist activities have killed nearly 350,000 people in North-East Nigeria and rendered over 310,000 refugees in internally displaced persons (IDP) camps in Nigeria (Global Conflict Tracker, 2021). Boko Haram terrorist group had claimed responsibility for several terror attacks, including the kidnapping of 276 female students in Borno state in 2014 (Varrella, 2021). Again, in 2020, 70 civilians were killed and several others injured in Borno state due to terrorist attacks (GOV.UK, 2020). It is shocking to note that Nigeria ranks third in the most terrorized countries in the world (Institute for Economics and Peace, 2020). These setbacks in Nigeria's growth and stability necessitate identifying and curbing terrorism.

Recently, associative mining has been used to perform the analysis, correlations, associations, and insights on different terrorist datasets to curb terrorist attacks around the globe (Khalifa, Taha, Taha and Hassanien, 2019; Li, Li, Tang, and Zhang, 2020; Atsa'am, Wario and Alola, 2021). For example, Khalifa, Taha, Taha and Hassanien (2019) made statistical insights and associative mining for terrorist attacks in Egypt by extracting and reviewing patterns of terrorist attacks in Egypt from the Global Terrorism Database (GTD). Motivated by Khalifa, Taha, Taha and Hassanien (2019), the researchers investigated the GTD to have clear insights into the terrorist activities in Nigeria.

In this paper, statistical derivations are first made to draw out pictorial meanings from the GTD datasets for Nigeria. Furthermore, the Apriori algorithm is used to develop an associative mining model to curb terrorist attacks in Nigeria using the GTD. The associative mining model aid insights into terrorist activities by making correlations and relationships of the numerous terrorist activities from 1970 to 2019 in Nigeria. The algorithm formulates rules that will serve as red flags in the lead-up to terrorist attacks while examining closely the frequency or manner in which terrorist activities occur across the States and regions of Nigeria. More so, findings will aid law enforcement agencies in effectively handling red flags to avert the loss of lives and properties due to terrorist attacks. The rest of the paper is organized as follows. Related works are described in Section 2. Section 3 describes our experiment, datasets used, and research questions. Results and discussions are presented in Section 4. Conclusion and future work are presented in Section 5.

## 2.0 RELATED WORKS

The Apriori algorithm was first introduced by Agrawal and Srikant (Agrawal, and Srikant 1994). Since then, it has been applied to different disciplines such as health care (Duru, 2005; Vashisht, Holy, Shah, Elangovanraaj, Johnston, and Coplan, 2020), flood areas prediction and flood risk analysis (Harun, Makhtar, Abd Aziz, Zakaria, and Syed, 2017; Zhong, Wang, Jiang, Huang, Chen, and Hong, 2020), cyberbullying analysis on social media (Zainol, Wani, Nohuddin, Noormanshah, and Marzukhi 2018), human behavior analysis (Raihan, Islam, Ghosh, Hassan, Angon, and Kabiraj, 2020), and the effective detection of factors that causes accidents (Nafie Ali, and Mohamed Hamed, 2018) amongst several other applications.

Considering terrorist attacks investigation, Singh, Chaudhary and Kaur, (2019) evaluated the terrorist activities from 1970 to 2017 on GTD. The researchers compared the performance of Gaussian Naïve Bayes, Linear Discriminant Analysis, K-Nearest Neighbour, Support Vector Machine, Logistic Regression, and Decision Tree algorithms. They showed that the Logistic regression, Linear Discriminant Analysis, Gaussian Naïve Bayes, and Support Vector Machine all gave prediction accuracies of 82%. Ghalleb, and Amara (2020) predicted terrorist attacks on the GTD and ACLED datasets using Support Vector Machine, Decision Tree, Random Forest and K-Nearest Neighbour algorithms using Tunisia as a case study. They showed that these four algorithms could effectively assist in curbing terrorist activities. Li, Li, Tang, and Zhang (2020) investigated the global cyberspace security issues using the Apriori algorithm by comparing the frequencies of words from global professional target websites. They showed that security issues in global cyberspace could be investigated and solutions provided via the use of the Apriori algorithm. Sathyavani (2021) used the Random Forest Algorithm, Gaussian Naive Bayes and Decision Tree algorithms on the GTD datasets to predict fear-based oppression in India. The researcher showed that the Random Forest algorithm achieved the highest fear-based oppression prediction accuracy of 96%. Atsa'am, Wario and Alola (2021) investigated and explained why terrorists attacked other terrorists using the Apriori algorithm on the GTD datasets and explained how these results could assist counter-terrorism. Iorliam, Dugeri, Akumba, and Otor (2021) utilised the Apriori algorithm as a forensic investigation tool to assist law enforcement agencies in making decisions when handling terrorist attacks/cases in Nigeria.

## 3.0 EXPERIMENTS

The first goal of this experiment is to investigate the Global Terrorism Database (GTD) to study and make deductions from the activities of terrorism in Nigeria from 1970-2019. The second goal of this experiment is to have an insight into terrorist activities in Nigeria based on the Apriori algorithm rules generated from the GTD. We use the Apriori algorithm developed by Agrawal and Srikant (1994) for this research because it is the most efficient and commonly used technique in associating how objects are strongly or weakly linked together. The Apriori algorithm is deployed to get patterns and insights about terrorism and terrorist activities in Nigeria. The Apriori algorithm model determines frequent patterns and associations based on the association rules using the GTD with a focus on Nigeria.

The Apriori rules are set in Python based on parameters such as observations ('iyear', 'imonth', 'extended', 'provstate', 'multiple', 'success', 'suicide', 'attacktype1_txt', 'targtype1 _txt', 'targsubtype1_txt'), the minimum support of the relation (min_support) in our case set to 0.1, the minimum confidence of relations (min_confidence) in our case set to 0.8, the minimum lift of relations (min_lift) in our case set to 0.01, and the minimum number of items in our rule (min_length) in our case set to 1. These observations are properly described in Section 3.1. Based on the above parameters, the association rules are generated, and the three important factors used to evaluate the rules generated in this investigation are Support, Confidence, and Lift.

For example, if we have $X \Rightarrow Y$, the three important factors are described as follows:

i. Support: This shows the popularity of an item in the data under consideration. Mathematically, it is expressed as:

   Support (X) = (Number of transactions containing (X)) / Total number of transactions.

ii. Confidence: This shows the probability of Y happening if X has happened. Mathematically, it is expressed as:

Confidence (X ⇒ Y) = (Number of Transactions containing both (X and Y)/ Total number of Transactions containing (X))

iii. Lift: This shows the ratio between Confidence and Support. It is expressed mathematically as:

Lift (X ⇒ Y) = Confidence (X ⇒ Y)/ Support (X)

In essence, to achieve top rules that we can make deductions and insights from the GTD with a focus on Nigeria, we expect a higher Support closer to 1 (100%), Confidence closer to 1 (100%), and Lift usually greater than 1.

### 3.1 Datasets

The National Consortium puts together the Global Terrorism Database (GTD) for the Study of Terrorism and Responses to Terrorism (START) (LaFree, and Dugan, 2007) available at: https://www.start.umd.edu/gtd was used to evaluate the proposed method in this paper. It comprises terrorist activities ranging from the year 1970 to 2019. This dataset has 135 columns, and the key nine features of the dataset used in this paper are described below:

1. iyear: this indicates the year an incident occurred or started
2. imonth: this indicates the month an incident occurred or started
3. extended: shows "Yes" when the duration of an incident is more than 24 hours, or as "No" otherwise
4. provstate: this shows the subnational administrative region where the incident occurred
5. multiple: when an attack is part of multiple incidents, it is reported as "Yes", or "No" otherwise
6. success: when an incident is successful, it is reported as "Yes", or "No" otherwise
7. suicide: when an incident is a suicide, it is reported as "Yes", or "No" otherwise
8. attacktype1/attacktype1_txt: this assigns a code (number) to a text (attack type hierarchy). For example, 1-the attacks classified in this category include Assassination, 2-Hijacking, 3-Kidnapping, 4-Barricade Incident, 5-Bombing/Explosion, 6-Armed Assault, 7-Unarmed Assault, 8-Facility/Infrastructure Attack, and 9-Unknown

9. targtype1/targtype1_txt: this assigns a code (number) to a text (general type of target/victim). For example, 1-Business, 2-Government (General), 3- Police, 4-Military, 5-Abortion Related, 6- Airports & Aircraft, 7-Government (Diplomatic), 8-Educational Institution, 9-Food or Water Supply, 10-Journalists & Media, 11- Maritime (Includes ports and maritime facilities), 12- NGO, 13-Other, 14-Private Citizens & Property, 15-Religious Figures/Institutions, 16-Telecommunications, 17- Terrorists/Non-State Militias, 18-Tourists, 19-Transportation (Other than Aviation), 20-Unknown, 21-Utilities, 22-Violent Political Parties.

The description of the other fields in the datasets is explained in LaFree, and Dugan (2007) and the Codebook: Inclusion Criteria and Variables (2019).

## 3.2 Research Questions

In this paper, we ask three key research questions that need to be answered:

1. What are the targets of these terrorist attacks?
2. Do attacks in Nigeria last for more than 24 hours (Extended)?
3. Are attacks in Nigeria usually successful?

# 4.0 RESULTS AND DISCUSSIONS

The results from these experiments are presented and explained in two sections, as shown in sections 4.1 and 4.2.

## 4.1 Deductions from the Global Terrorism Database from 1970-2019

Reading the GTD into the Python programming environment and extracting the data of interest, which focuses on Nigeria, the relevant figures are achieved. Figure 1 presents the number of attacks that have occurred in Nigeria from 1970-2019.

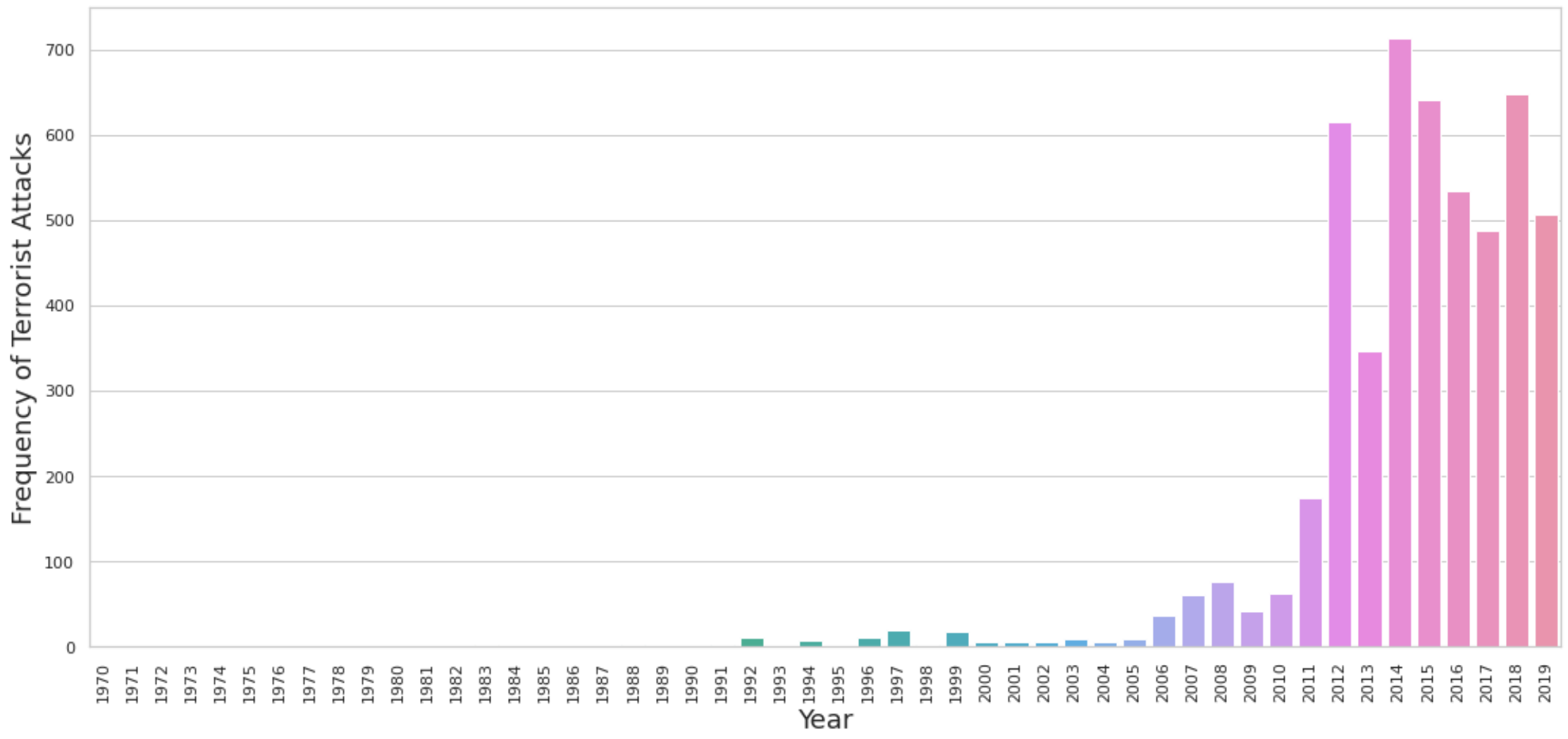

Figure 1: Frequency of Terrorist Attacks VS Years in Nigeria

As shown in Figure 1, the number of attacks occurred more between the years 2011 to 2019. As reported by Global Conflict Tracker (2021), the Boko Haram insurgence erupted in 2011, and as such, the level of terrorist attacks increased. Specifically, in August 2011, Boko Haram bombed the United Nations house in Abuja, killing 23 persons. From 1970 to 2019, the peak of these attacks were in 2012 (12.1%), 2014 (14.1%), 2015 (12.6%), and 2018 (12.8%). The highest frequency of attacks in 2014 (14.1%), as shown in Figure 1 was also described by reliefweb (2015) that Nigeria had 7,512 fatalities, increasing terrorist attacks by 300% and rating Nigeria the third most terrorized country in the world behind Afghanistan and Iraq (reliefweb, 2015). It means that the frequency of terrorist attacks would not have increased if not for the emergence and operations of Boko Haram in Nigeria. As at 2014, Boko Haram was rated to be the most deadly terrorist group more than ISIL, Taliban, and al-Qa'ida (Institute for Economics and Peace, 2016). Furthermore, an in-depth look at the target areas these terrorist attacks occurred more frequently is shown in Figure 2.

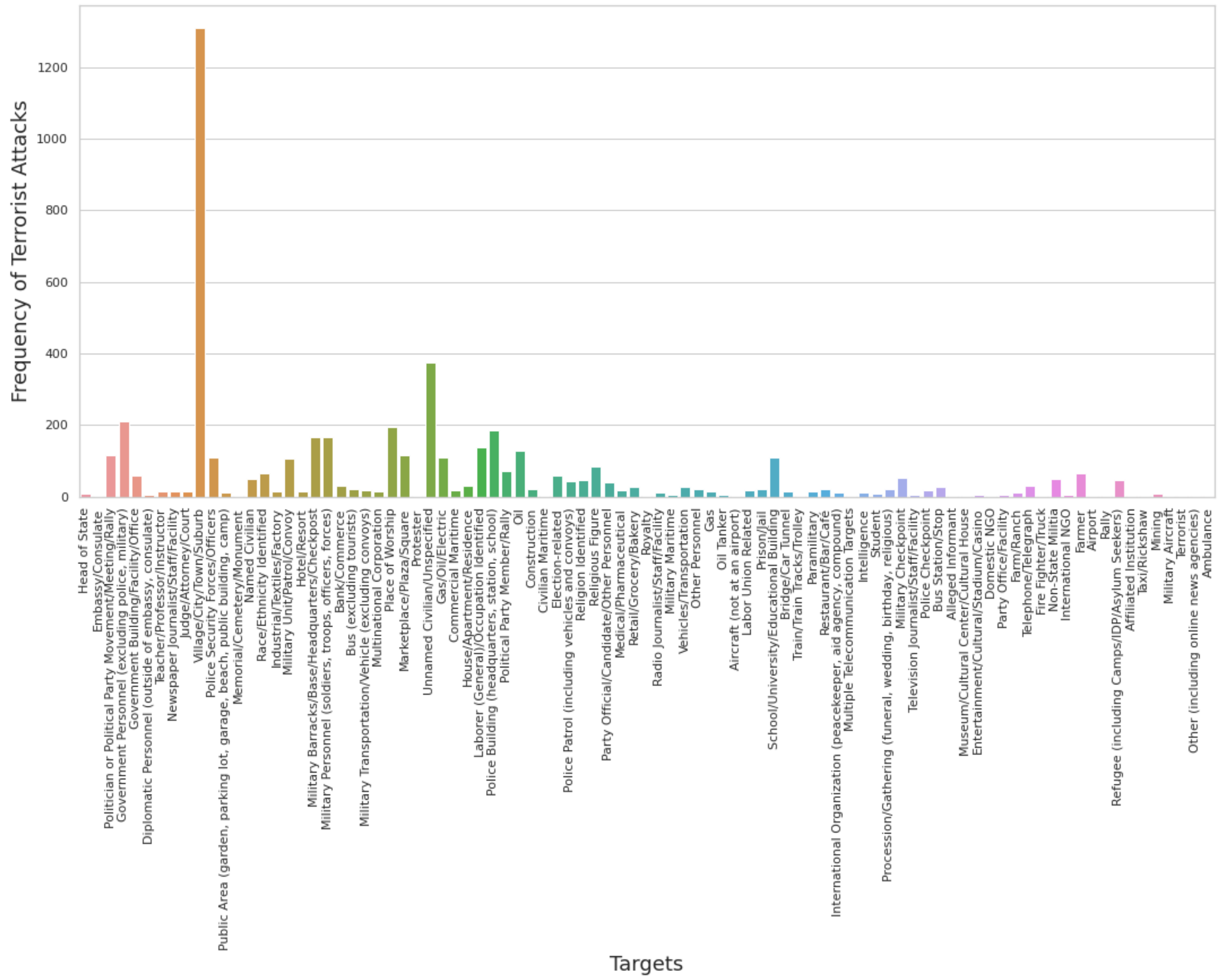

Figure 2: Frequency of Terrorist Attacks VS Targets

From Figure 2, the top three targets with the highest frequency of attacks are the Village/City/Town/Suburb (26.6%), followed by Unarmed Civilians/Unspecified (7.6%), and Government Personnel (excluding police, military) (4.3%). This insight agrees with Sales (2019), where the research showed that terrorists usually target civilians in villages of Nigeria. This means that the Nigerian government needs to improve on securing the villages either by using the Vigilante Group of Nigeria or forming community police to safeguard lives and properties. More importantly, the government and law enforcement agencies need to provide digital surveillance to cover not only the City, Town or Suburb but also the villages to secure lives and properties.

Furthermore, to have an insight into attacks that are successfully launched in Nigeria from 1970 to 2019, Figure 3 shows the pie chart.

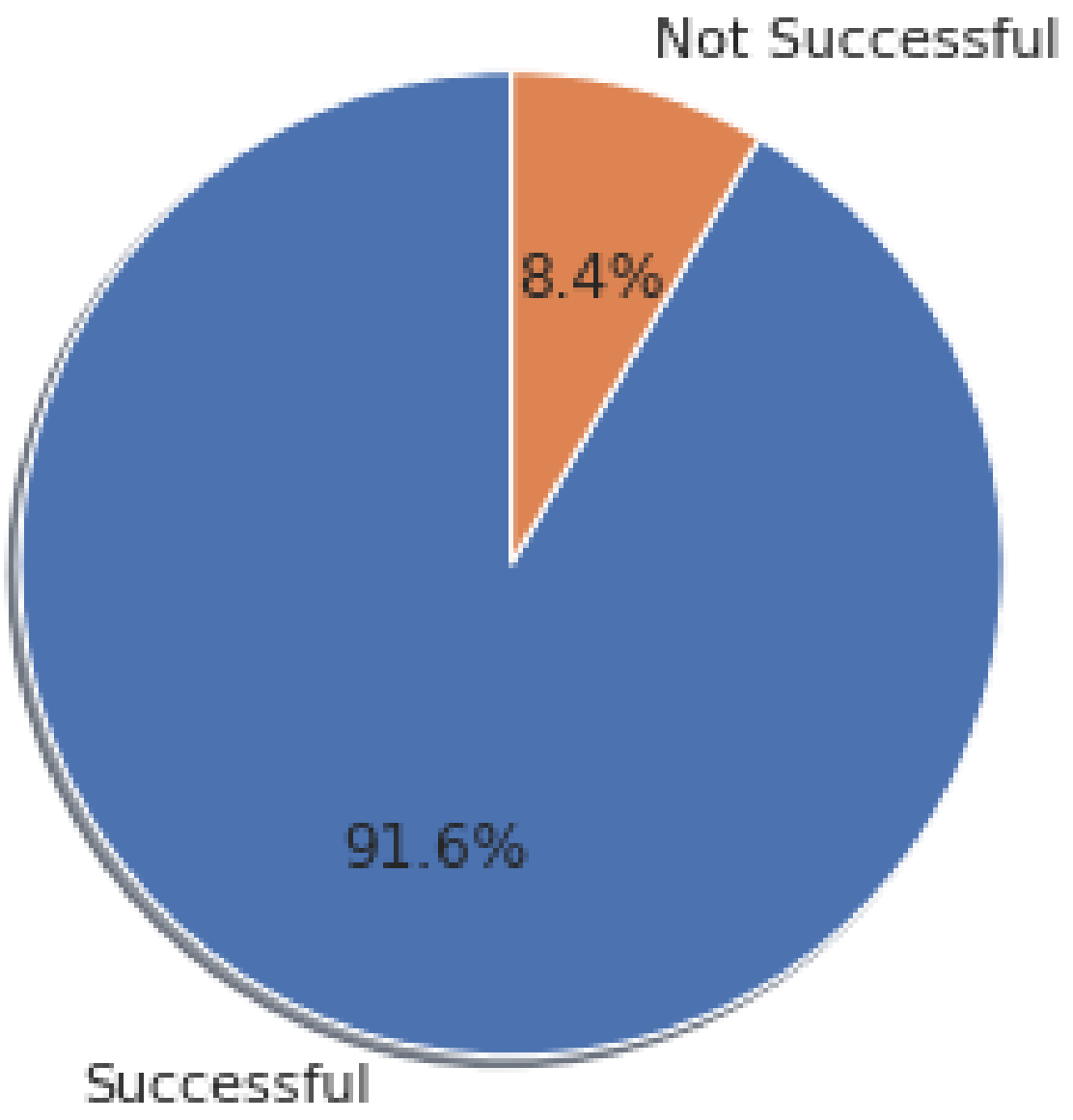

Figure 3: Successful VS Not Successful Attacks in Nigeria 1970-2019

It is observed from Figure 3 that 91.6% of attacks launched in Nigeria were all successful. It is very scary, and the Nigerian government and law enforcement agencies need to be very proactive in curbing terrorism, or else, soon Nigeria will be inhabitable by its citizens.

Also, Figure 4 shows the ratio of suicidal attacks (weapon carried bodily by human being) and non-suicidal attacks (attack carried out by using other weapon types and not carried by a human being) in Nigeria from 1970 to 2019.

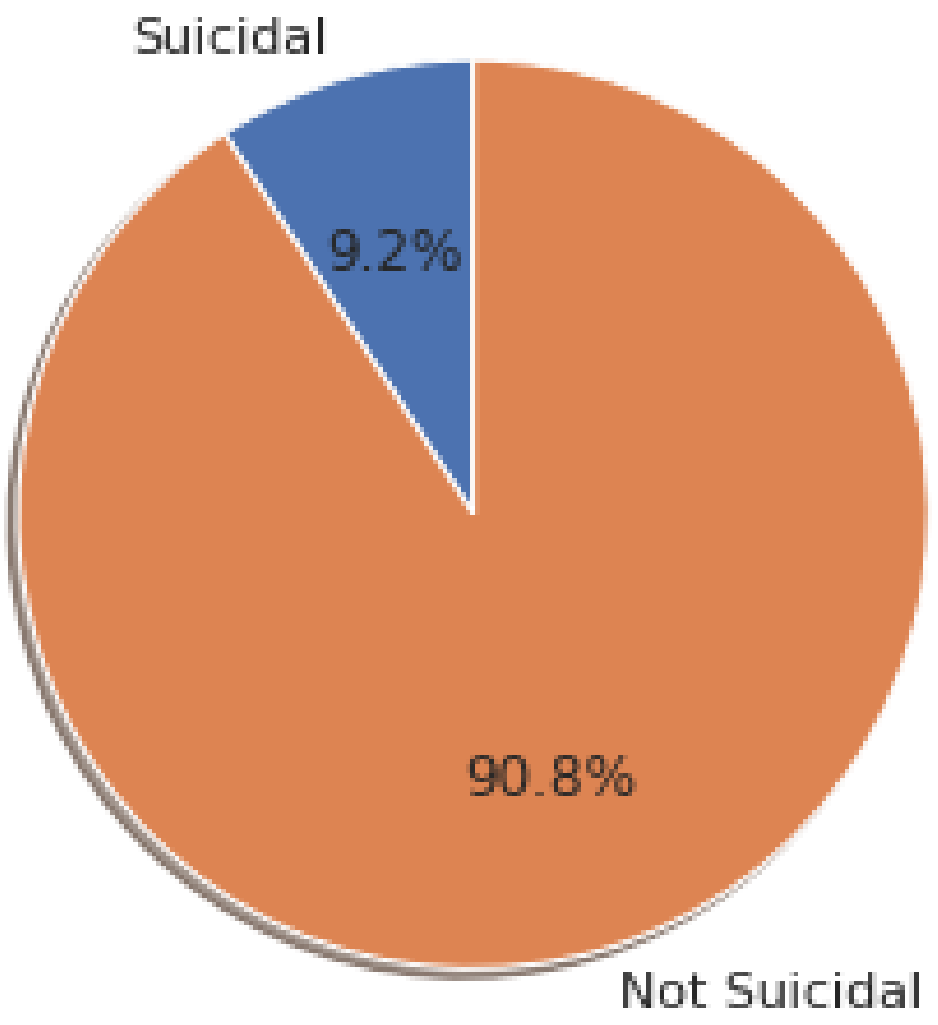

Figure 4: Suicidal Attacks VS Non Suicidal Attacks in Nigeria 1970-2019

It can be observed from Figure 4 that in Nigeria, 90.8% of the attacks are usually non suicidal and only 9.2% of the attacks are suicidal. The implication of this is that most terrorists in Nigeria prefer not to die during their attacks.

Again, Figure 5 shows the heat map indicating states that are more prone to attacks from 1970 to 2019.

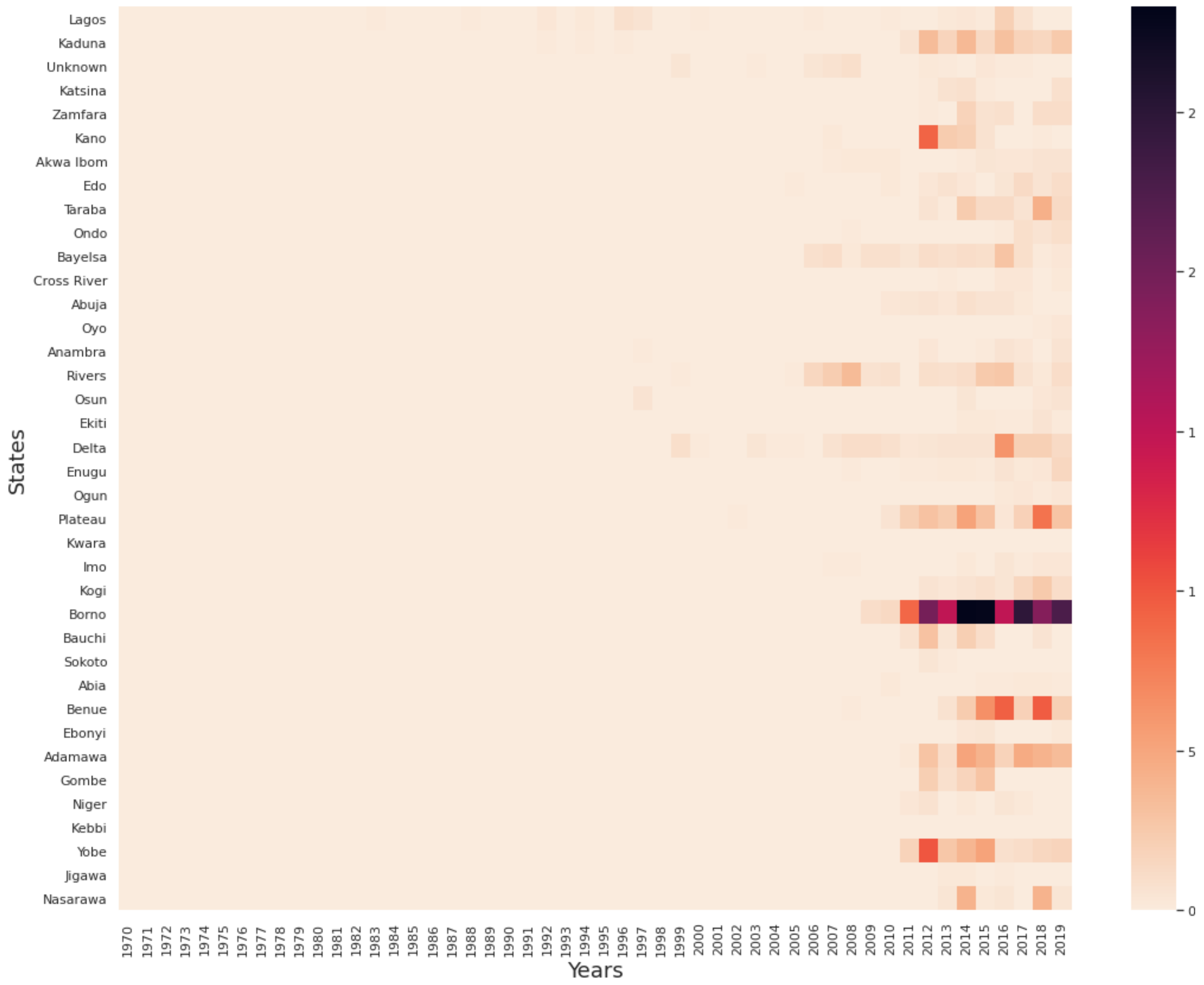

Figure 5: Heat Map of Attacks in Nigeria 1970-2019

It is observed in Figure 5 that terrorist attacks frequently target Borno State, Adamawa State, Yobe State, Benue State, Plateau State, Rivers State, Kaduna State, Plateau State, and Kano State. It is interesting to note that Kwara State, Sokoto State, Eboni State, Kebbi State, and Jigawa State had little or no attacks from 1970 to 2019. It shows that investors and tourists should feel safer in these states compared to other states affected by terrorist attacks in Nigeria. Even though there is a need to protect lives and properties in Nigeria, law enforcement agencies such as the Nigeria Police should be deployed more to the most affected states of Nigeria.

### 4.2 Deductions Based on the Apriori Algorithm Rules Generated from the Global Terrorism Database from 1970-2019

The Apriori algorithm described in the methodology generated 214 strong rules and 49 out of the 214 strong rules are presented in Table 1. These 49 strong rules are selected and analysed due to their high Confidence, Lift, and relevance in answering the research questions posed in Section 3.2.

Table 1: Strong Rules Based on Confidence and Lift

| No. | Left Hand Side | Right Hand Side | Support | Confidence | Lift |
|---|---|---|---|---|---|
| 1 | 2012 ⇒ | Not Extended | 0.11715976331360947 | 0.9658536585365854 | 1.1114112684476825 |
| 2 | 2012⇒ | Not Suicide | 0.11420118343195267 | 0.9414634146341464 | 1.0369801243091727 |
| 3 | 2014⇒ | Not Extended | 0.12347140039447732 | 0.8779803646563815 | 1.0102951540644245 |
| 4 | 2014⇒ | Not Suicide | 0.13274161735700196 | 0.943899018232819 | 1.0396628334652167 |
| 5 | 2014⇒ | Successful | 0.13175542406311638 | 0.9368863955119215 | 1.022828170810819 |
| 6 | 2015⇒ | Not Extended | 0.1136094674556213 | 0.8985959438377535 | 1.0340175749562892 |
| 7 | 2018⇒ | Successful | 0.11775147928994083 | 0.9212962962962964 | 1.0058079720547421 |
| 8 | Not Extended⇒ | Armed Assault | 0.4287968441814596 | 0.9362618432385875 | 1.0773598604674623 |
| 9 | Not Suicide ⇒ | Armed Assault | 0.45779092702169627 | 0.9995693367786392 | 1.10098121604773 |
| 10 | Successful⇒ | Armed Assault | 0.4416173570019724 | 0.9642549526270457 | 1.0527072803228084 |
| 11 | Not Extended ⇒ | Bombing/Explosion | 0.2493096646942801 | 0.9976322020520916 | 1.1479789524294381 |
| 12 | Extended⇒ | Not Suicide | 0.13037475345167654 | 0.9954819277108435 | 1.0964791165531125 |
| 13 | Extended⇒ | Successful | 0.13076923076923078 | 0.9984939759036146 | 1.0900870925562718 |
| 14 | Hostage Taking (Kidnapping) ⇒ | Not Suicide | 0.12248520710059171 | 0.9951923076923076 | 1.0961601129698022 |
| 15 | Successful⇒ | Hostage Taking | 0.11972386587771203 | 0.9727564102564101 | 1.0619885874246338 |

| No. | Left Hand Side | Right Hand Side | Support | Confidence | Lift |
|---|---|---|---|---|---|
| | | (Kidnapping) | | | |
| 16 | Multiple⇒ | Not Extended | 0.31005917159763313 | 0.9107763615295481 | 1.0480336252734475 |
| 17 | Not Suicide ⇒ | Not Multiple | 0.6051282051282051 | 0.9174641148325359 | 1.0105459618077248 |
| 18 | Successful⇒ | Not Multiple | 0.5879684418145956 | 0.8914473684210527 | 0.9732209642322862 |
| 19 | Private Citizens & Property⇒ | Village/City/Town/Suburb | 0.2579881656804734 | 1.0 | 2.1133805752396833 |
| 20 | 2012⇒ | Successful | 0.10552268244575937 | 0.8699186991869919 | 1.1078844021296281 |
| 21 | Successful ⇒ | 2015 | 0.10236686390532544 | 0.8096723868954758 | 1.0311577497010957 |
| 22 | 2018⇒ | Not Suicide | 0.10808678500986194 | 0.8456790123456791 | 1.0095579450418162 |
| 23 | Borno⇒ | Not Suicide | 0.1601577909270217 | 0.9987699876998771 | 1.1001007685505926 |
| 24 | Not Extended⇒ | Multiple | 0.16351084812623273 | 0.9089912280701754 | 1.0459794658002246 |
| 25 | Successful⇒ | Multiple | 0.17692307692307693 | 0.9835526315789475 | 1.0737751597987217 |
| 26 | Not Multiple⇒ | Not Extended | 0.26528599605522685 | 0.9539007092198584 | 1.0976569668054204 |
| 27 | Not Extended⇒ | Not Suicide | 0.42859960552268245 | 0.9358311800172265 | 1.2036184887588377 |
| 28 | Private Citizens & Property⇒ | Not Extended | 0.2562130177514793 | 0.9258731290092659 | 1.0654055297496545 |
| 29 | Successful⇒ | Not Extended | 0.4124260355029586 | 0.9005167958656332 | 1.1468525885553278 |
| 30 | Village/City/Town/Suburb⇒ | Not Extended | 0.16587771203155818 | 0.90625 | 1.0428251248297777 |
| 31 | Not Multiple⇒ | Not Suicide | 0.2781065088757396 | 1.0 | 1.1014555724527482 |
| 32 | Not Multiple⇒ | Successful | 0.26469428007889545 | 0.9517730496453901 | 1.0390803965766857 |
| 33 | Private Citizens & Property⇒ | Not Suicide | 0.2767258382642998 | 1.0 | 1.1014555724527482 |
| 34 | Successful⇒ | Not Suicide | 0.4414201183431953 | 0.9638242894056849 | 1.1505978684452136 |
| 35 | Village/City/Town/Suburb⇒ | Not Suicide | 0.18303747534516765 | 1.0 | 1.1014555724527482 |
| 36 | Successful⇒ | Village/City/Town/Suburb | 0.17810650887573964 | 0.9730603448275862 | 1.0623204022988506 |
| 37 | Borno⇒ | Not Extended | 0.11617357001972387 | 0.9966159052453469 | 1.1468094960494575 |
| 38 | Not Multiple⇒ | Not Extended | 0.15976331360946747 | 0.9963099630996312 | 1.1464574473252678 |

| No. | Left Hand Side | Right Hand Side | Support | Confidence | Lift |
|---|---|---|---|---|---|
| 39 | Not Extended ⇒ | Not Suicide | 0.15897435897435896 | 0.9962917181705809 | 1.146436452819983 |
| 40 | Borno⇒ | Multiple | 0.11577909270216963 | 0.9100775193798449 | 1.0472294651057226 |
| 41 | Borno⇒ | Private Citizens & Property | 0.16962524654832348 | 0.9502762430939227 | 1.0374462860650706 |
| 42 | Successful⇒ | Hostage Taking (Kidnapping) | 0.10019723865877712 | 0.8141025641025641 | 1.3846024823884604 |
| 43 | Private Citizens & Property⇒ | Multiple | 0.1940828402366864 | 0.9283018867924528 | 1.0224832861259474 |
| 44 | Village/City/Town/Suburb⇒ | Multiple | 0.14911242603550295 | 0.9921259842519684 | 1.0927826939294982 |
| 45 | Village/City/Town/Suburb⇒ | Successful | 0.14694280078895464 | 0.9776902887139107 | 1.0673750568000706 |
| 46 | Borno⇒ | Not Multiple | 0.10493096646942801 | 1.0 | 1.1014555724527482 |
| 47 | Borno⇒ | Successful | 0.15325443786982249 | 0.955719557195572 | 1.1409225700451024 |
| 48 | Armed Assault⇒ | Private Citizens & Property | 0.11183431952662722 | 1.0 | 2.2684563758389262 |
| 49 | Not Extended⇒ | Armed Assault | 0.13076923076923078 | 0.9567099567099567 | 1.2304717099237648 |

From Table 1, six top strong rules with the highest confidence (1) are “Private Citizens & Property ⇒ Village/City/Town/Suburb, Not Multiple ⇒ Not Suicide, Private Citizens & Property ⇒ Not Suicide, Village/City/Town/Suburb ⇒ Not Suicide, Borno ⇒ Not Multiple, and Armed Assault ⇒ Private Citizens & Property.”

The first top strong rule “Private Citizens & Property ⇒ Village/City/Town/Suburb” shows that the probability of an attack occurring in the “Village/City/Town/Suburb” and on “Private Citizens & Property” is 1 (100%) which is very high. Furthermore, this rule has a high Lift of 2.11338 which shows that there exists a positive correlation between “Private Citizens & Property” and “Village/City/Town/Suburb”. Moreover, the frequency of an attack occurring in the “Village/City/Town/Suburb” and on “Private Citizens & Property” is 25.7988%. This shows that terrorist perform their attacks more on private citizens and properties. Again, it could be deduced that these attacks occur more in the villages, cities, towns and suburbs. This technically means that terrorist attacks are not only occurring in villages but these attacks have spread to cities, towns and suburbs.

The second top strong rule “Not Multiple ⇒ Not Suicide” shows that the probability of an attack been “Not suicide” and “Not Multiple” is 1 (100%) which is very high. Moreover, a Lift of 1.1015 shows that there exists a positive correlation between attacks that are “Not Suicide” and “Not Multiple”. Again, the frequency of an attack been “Not suicide” and “Not Multiple” is 27.81065%. We can infer from this rule that attacks in Nigeria are not due to terrorists carrying out the attacks and in the cause, losing their lives (suicide), but primarily due to terrorists performing these attacks and also preserving their lives.

The third top strong rule “Private Citizens & Property ⇒ Not Suicide” shows that the probability of an attack been “Not suicide” and occurring on “Private Citizens & Property” is 1 (100%) which is very high. The third rule has a Lift of 1.1015 which shows a positive correlation between “Private Citizens & Property” and “Not suicide”. Moreover, the frequency of an attack been “Not suicide” and occurring on “Private Citizens & Property” is 27.67258%. It could be inferred again that private citizens and properties in Nigeria are the targets of terrorist attacks.

Again, the fourth top strong rule “Village/City/Town/Suburb ⇒ Not Suicide” shows that the probability of an attack been “Not suicide” and in the “Village/City/Town/Suburb” is 1 (100%) which is very high. Furthermore, the fourth rule has a Lift of 1.1015 which shows that there exists a positive correlation between “Not suicide” and “Village/City/Town/Suburb”. Also, the frequency of an attack been “Not suicide” and occurring in the “Village/City/Town/Suburb” is 18.3037%. This agrees to the fact that terrorist attacks occur more in the villages, cities, towns and suburb in Nigeria.

The fifth top strong rule “Borno ⇒ Not Multiple” shows that the probability of an attack been “Not Multiple” and occurring in “Borno” is 1 (100%) which is very high. Moreover, the fifth rule has a Lift of 1.014556 which shows that there exists a positive correlation between “Not Multiple” and “Borno”. Furthermore, the frequency of an attack been “Not Multiple” and occurring in “Borno” is 10.4930966%. Even though every state in Nigeria needs maximum security, Borno state has the highest number of attacks based on the confidence of 1 and Lift of 1.014556. Hence, law enforcement agencies and the Nigerian government should improve on physical security and digital surveillance in Borno state.

The sixth top strong rule “Armed Assault ⇒ Private Citizens & Property” shows that the probability of an attack type been an “Armed Assault” and occurring on “Private Citizens & Property” 1 (100%) which is very high. The sixth rule has a lift of 1.2305 which shows a very high correlation between “Armed Assault” and “Private Citizens & Property”. Moreover, the frequency of an attack type been “Armed Assault” and occurring on “Private Citizens & Property” is 11.18343%. This means that guns and weapons are usually used by terrorist attackers to destroy lives and properties in Nigeria. It could be inferred that there could be a free influx of weaponry in Nigeria to grant terrorists access to these weapons. Another reason for this could be that there exists a high chance of several locally manufactured weapons in Nigeria. The law enforcement agencies and the Nigerian government needs to curb the influx of these weapons to weaken/curb the activities of terrorists in Nigeria. Moreover, law enforcement agencies need to perform more investigations in discovering illegal weapon manufacturing companies/homes.

To further answer the research questions raised, the rules that contain “Not Extended” on the RHS shows that attacks were “Not Extended” in “2012”, attacks were “Not Extended” in “2014”, attacks were “Not Extended” in “2015”, attacks were “Not Extended” and “Not Multiple”, attacks were “Not Extended” on “Private Citizens & Property”, attacks were “Not Extended” and “Successful”, attacks were “Not Extended” in “Village/City/Town/Suburb”, attacks were “Not Extended” in “Borno”, and attacks were “Not Extended” and “Not Multiple”. Based on this rule, we can infer that terrorist attacks in some parts of Nigeria are yet to extend for more than 24 hours. However, law enforcement agencies and the Nigerian government should act quickly to avoid attacks getting to the level of extending for more than 24 hours in such areas.

Furthermore, results from the experiment with the rule “Successful” on the RHS shows that attacks in Nigeria were “Successful” in “2014”, attacks were “Successful” in “2018”, attacks were “Successful” and “Extended”, attacks were “Successful” in “2012”, attacks were “Successful” and “Not Multiple”, attacks were “Successful” in the “Village/City/Town/Suburb”, and attacks were “Successful” in “Borno”. It can be inferred from this rule that most of the attacks in Nigeria have been successful. Again, it is noted that some attacks that were successful also occurred longer than 24 hours (Extended). This means that law enforcement agencies and the government need to act very fast in curbing these attacks.

Again, from the experiment, the rules that contain “Borno” on the LHS shows that attacks that are “Not Suicide” occurs in “Borno”, attacks that are “Not Extended” occur in “Borno”, attacks that are “Multiple” occur in “Borno”, attacks that are on “Private Citizens & Property” happen in “Borno”, attacks that are “Not Multiple” happen in “Borno”, and attacks that are “Successful” happen in “Borno”. From this rule, it is clear that attacks appear to occur more frequently in Borno state as compared to other states. This does not in any way mean that other states are not affected by terrorist attacks. Hence, law enforcement agencies and the government needs to concentrate more efforts in curbing terrorist attacks in Borno state and other affected states.

## 5.0 CONCLUSION AND FUTURE WORK

Terrorism has negatively affected the entire world, and Nigeria is not excluded. In this paper, we investigated and made insights into terrorist activities in Nigeria. From the experiments, we observed that the number of attacks increased tremendously from 2011 to 2019. Moreover, the target with the highest frequency of attacks is the Village/City/Town/Suburb. Again, our experiment supported our conjecture that attacks in Nigeria are mostly successful. The main application of this work is to assist law enforcement agencies with the needed information to be one step ahead of terrorists in making the right decisions targeted at curbing terrorism in Nigeria. In our future work, we plan to use Power laws such as Benford’s law and Zipf’s law in the investigation of terrorist attacks in Nigeria. Moreover, the behavior associations in lone actor terrorists (Altay, Baykal-Gürsoy, and Hemmer, 2020) will be investigated in the future using the Power laws.